\documentclass[aip,graphicx,amsmath,amssymb,reprint]{revtex4-1}
\usepackage{graphicx}
\usepackage{dcolumn}
\usepackage{bm}
\usepackage[utf8]{inputenc}
\usepackage[T1]{fontenc}
\usepackage{mathptmx}
\usepackage{etoolbox}
\usepackage[]{units}
\usepackage{xcolor}

\DeclareUnicodeCharacter{0308}{\"}
\DeclareUnicodeCharacter{030C}{\v{ }}
\DeclareUnicodeCharacter{0301}{\'{ }}

\newcommand{\RomanNumeralCaps}[1]{\MakeUppercase{\romannumeral #1}}
\usepackage{titlesec}
\titleformat{\section}
  {\normalfont\large\bfseries\raggedright}{\thesection.}{0.5em}{}
\titlespacing*{\section}
  {0pt}{3.5ex plus 1ex minus .2ex}{1.5ex plus .2ex}

\titleformat{\subsection}
  {\normalfont\normalsize\bfseries\raggedright}{\thesubsection.}{0.5em}{}
\titlespacing*{\subsection}
  {0pt}{2.5ex plus 1ex minus .2ex}{1ex plus .2ex}
\date{\Today}

\makeatletter
\def\@email#1#2{%
 \endgroup
 \patchcmd{\titleblock@produce}
  {\frontmatter@RRAPformat}
  {\frontmatter@RRAPformat{\produce@RRAP{*#1\href{mailto:#2}{#2}}}\frontmatter@RRAPformat}
  {}{}
}%
\makeatother
\draft 

\begin{document}


\title{Hybrid magnonic–spintronic system for tunable broadband signal filtering and microwave generation}



\author{A. Koujok}
\affiliation{Fachbereich Physik and Landesforschungszentrum OPTIMAS, Rheinland-Pf\"alzische Technische Universit\"at Kaiserslautern-Landau, 67663 Kaiserslautern, Germany}
\author{A. Hamadeh*} 
\affiliation{Centre de Nanosciences et de Nanotechnologies, CNRS, Universit\'e Paris-Saclay, 91120, Palaiseau, France}
\email{alexandre.hamadeh@universite-paris-saclay.fr}
\author{L.Martins} \affiliation{Univ. Grenoble Alpes, CEA, CNRS, GINP, Spintec, Grenoble, France}
\author{F. Kohl} \affiliation{Fachbereich Physik and Landesforschungszentrum OPTIMAS, Rheinland-Pf\"alzische Technische Universit\"at Kaiserslautern-Landau, 67663 Kaiserslautern, Germany}
\author{B. Heinz} \affiliation{Fachbereich Physik and Landesforschungszentrum OPTIMAS, Rheinland-Pf\"alzische Technische Universit\"at Kaiserslautern-Landau, 67663 Kaiserslautern, Germany}
\author{U. Ebels} \affiliation{Univ. Grenoble Alpes, CEA, CNRS, GINP, Spintec, Grenoble, France}
\author{P. Pirro} \affiliation{Fachbereich Physik and Landesforschungszentrum OPTIMAS, Rheinland-Pf\"alzische Technische Universit\"at Kaiserslautern-Landau, 67663 Kaiserslautern, Germany}

\date{\today}




\begin{abstract}

Non-conventional beyond-the-state-of-the-art signal processing schemes require parallelism, scalability, robustness and energy efficiency to meet the demands of complex data-driven applications. With further research, magnonic and spintronic circuits can potentially help to fulfill these requirements. We present an experimental proof-of-concept of a hybrid device that can employ broad deteriorated microwave signals to excite and detect low energy propagating spin waves (SWs). For this, we use the output signal of a spin-transfer torque nano-oscillator (STNO) and connect it to a RF filter based on a magnonic delay-line. The STNO serves as a tunable nano-scaled signal generator with a broad output linewidth. Its RF output is fed as input into the magnonic delay-line circuit. Tuning the magnetic field solely at the magnonic circuit, we demonstrate the capability to selectively filter a broad RF input, obtaining a spin-wave output signal with a much narrower linewidth. This allows to tune the frequency of the RF signal at the output simply by tuning the magnetic field. Our findings are a first step towards a versatile, energy-efficient and compact wave-based filter with high sensitivity. Such a device can use even low-power, degraded signals and convert them into tunable SW outputs, effectively reducing the need for charge-based signal processing.

\end{abstract}


\maketitle 

The rise of complex data-driven applications such as artificial intelligence, has driven a surge in demand for unconventional signal processing methods that allow for parallelism while adhering to stringent energy constraints \cite{PilzMahmoodHeim2025,deRoucyRochegondeBuffard2025}. With the size of complementary metal-oxide-semiconductor (CMOS) transistors approaching the atomic scale, effects such as Joule heating and current leakage due to quantum tunneling become more critical as they can potentially hinder performance. Spin-wave (SW) based data processing is one promising route which can help extend the envelope of CMOS technology. Magnonic systems – which manipulate magnons (quanta of SWs) – exhibit intrinsic non-linearity, element of memory and cascadability \cite{dutta2015non,fischer2017experimental,csaba2017perspectives,mahmoud2023two}. The latter enable programmable and reconfigurable circuits, the reason these systems are pursued for applications in magnon computing \cite{klingler2015spin,chumak2022advances,nakane2023performance,namiki2024fast,gubbiotti20242025} and magnonic RF processing \cite{vogt2014realization,wang2018reconfigurable,erdelyi2025design,davidkova2025nanoscale,kohl2025identification}.


\begin{figure}[!ht]
\centering
\includegraphics[width=\columnwidth]{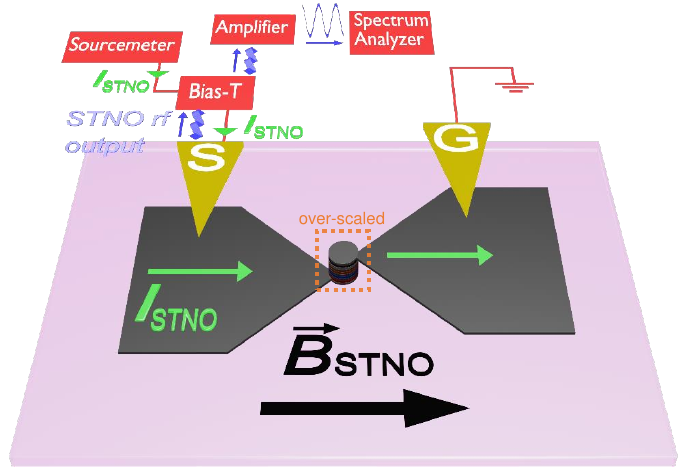}
\caption{Schematic of the circuit constituting the STNO layer. Dc current $I_{\mathrm{STNO}}$ is injected into the STNO via a bias-T, which in turn is used to direct back the RF signal generated by the STNO through an amplifier and onto a spectrum analyzer. The STNO is placed between the poles of an electric magnet that generates and in-plane magnetic field $B_{\mathrm{STNO}}$.}\label{fig1}
\end{figure}

Another promising route in the context of "beyond CMOS" technologies is the use of spintronic elements (those that aim at exploiting both the charge and the spin of an electron) to overcome or compensate some of the limitations of CMOS technology \cite{vzutic2004spintronics}.  One prominent spintronic device that is under intensive research and prototyping in this regard is the spin-transfer torque nano-oscillator (STNO) \cite{slavin2009nonlinear,sharma2017resonant}. STNOs are intrinsically non-linear devices that operate via electric currents based on the principle of Spin-transfer torque (STT) \cite{slonczewski1996current,berger1996emission,tsoi1998excitation,myers1999current}. Typical STNOs consist of a tri-layer stack whereby two ferromagnetic layers, namely the polarizing and the free layers, are separated by a non-magnetic spacer. Depending on whether the spacer is metallic or insulating, a STNO is referred to as a giant magneto-resistance valve \cite{baibich1988giant,binasch1989enhanced} or a magnetic tunnel junction (MTJ) \cite{julliere1975tunneling}, respectively. Injecting a dc current into the STNO leads to the spin-polarization of its electrons due to the polarizing layer. The spin-polarized current then transfers spin angular momentum onto the free layer, thus exerting torque on its magnetization. The precessing magnetization generates a time-varying magneto-resistance, leading to a microwave voltage signal \cite{slavin2009nonlinear,wada2010spin,hamadeh2024core,hamadeh2025diverse}. In the context of microwave generation, a STNO can be thought of as a micro-scaled tunable microwave generator, with typical generation powers lying in the sub-microwatt range and frequencies ranging from a few hundreds of \unit[]{MHz} to several tens of \unit[]{GHz} \cite{deac2008bias,dussaux2010large,hamadeh2022hybrid}.


Although still challenging, merging spintronic and magnonic approaches can open promising perspectives for hybrid devices capitalizing on strengths from both worlds. In this work, we propose to integrate a STNO with a magnonic circuit to realize a tunable, broadband signal filter. The presented approach is a fully experimental proof-of-concept for the device combining a STNO and a RF filter based on a magnonic delay line aiming to harness advantages from both spintronic and magnonic elements, namely microwave generation and wave-based circuitry, respectively. The STNO is used as a broadband nano-scaled microwave generator, whereby it outputs signal due to noise amplification induced by the injected current. That signal is detected by the SW system and selectively filtered by using the frequency (wave vector) dependent SW excitation efficiency. Then that excited SW signal is detected again at the output end of the circuit amplifying the filtering effect further. 

\begin{figure}[!ht]
\centering
\includegraphics[width=\columnwidth]{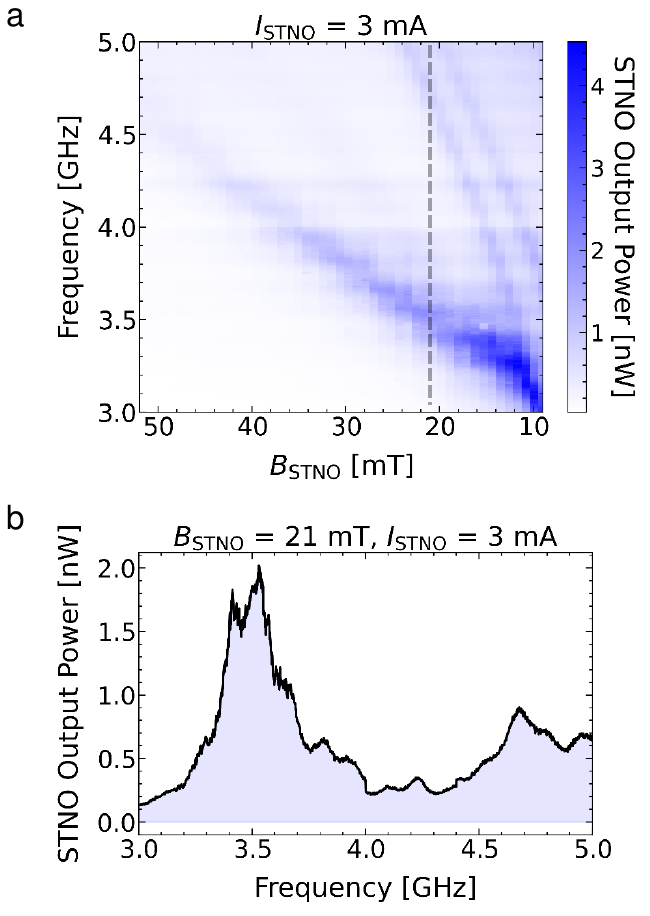}
\caption{(a) 2D color-plot of the frequency dependence of the STNO's output signal on the applied magnetic field $B_{\mathrm{STNO}}$ at $I_{\mathrm{STNO}}$ = \unit[3]{mA}. The dotted black line represents the chosen field at which the STNO will be operated throughout the experiment. (b) Power spectrum of the STNO's microwave output at fixed $B_{\mathrm{STNO}}$ and $I_{\mathrm{STNO}}$.}\label{fig2}
\end{figure}

To realize the feasibility of the proof-of-concept, the two circuits will be connected electrically. First steps, however, involve studying each circuit separately before the actual connection between both is carried out. In FIG. \ref{fig1}, we present the spintronic circuit comprising the connections to and from the STNO. The orange dots highlight the over-scaling of STNO with respect to the contact electrodes. The STNO consists of a multilayered stack, the diameter of which is \unit[300]{nm}. The stack's composition is as follows:\\ $IrMn/CoFe_{30}/Ru/CoFe_{40}B_{20}/MgO/CoFe_{40}B_{20}/Ta/NiFe_{19}$ of respective thicknesses:\\ $\unit[6]{nm}/\unit[2.6]{nm}/\unit[0.85]{nm}/\unit[1.8]{nm}/\unit[1]{nm}/\unit[2]{nm}/\unit[0.21]{nm}/\unit[7]{nm}$. The same stack has been previously used in \cite{phan2024unbiased}, where sputtering along with ion beam and optical etching techniques were employed to deposit and fabricate the material stack.


The STNO is placed in an electro-magnet that generates an in-plane magnetic field $B_{\mathrm{STNO}}$. A source-meter is used to supply the STNO with dc current $I_{\mathrm{STNO}}$. The dc current is directed through the dc port of a bias-T, and injected into the structure via a RF probe. The RF output is directed back through the RF port of the bias-T and detected by a spectrum analyzer after being passed through an amplifier (working range  \unit[3-8]{GHz} with an amplification of \unit[30]{dB}). First, the STNO's output signal is investigated as a function of $B_{\mathrm{STNO}}$ for fixed $I_{\mathrm{STNO}}$. FIG.~\ref{fig2}(a) shows the power spectrum as a function of field $B_{\mathrm{STNO}}$ and detection frequency. We fix $I_{\mathrm{STNO}}$ at \unit[3]{mA}, which provides a sufficiently large microwave output signal. The STNO's free layer is first saturated at high fields above \unit[500]{mT}. Thereafter, we sweep $B_{\mathrm{STNO}}$ from \unit[52]{mT} to \unit[9]{mT} at \unit[3]{mA}. The field at which the STNO is later operated throughout the experiment is marked with the dotted black line. Three modes are observed, two that increase their frequencies sharply with any field increase, and one whose frequency exhibits a steady linear increase within the swept $B_{\mathrm{STNO}}$ range. Based on their spectral distribution, these modes do not match the standard ones expected for such systems \cite{kim2012spin,urazhdin2014nanomagnonic}. The main mode at the lowest frequencies derives from noise-driven excitation of the uniform mode (FMR), while the other two may be either attributed to picked-up signals from the synthetic anti-ferromagnet, or to non-uniform modal excitations. These studied STNO stacks were designed specifically for experiments investigating vortex-based MTJs. The same device stacks had been studied operating in the vortex state in \cite{bendjeddou2023electrical}. Nevertheless, due to technical limitations imposed by the Yttrium-Iron-Garnet (YIG) delay line (see rationale in the section below), high frequencies in the \unit[]{GHz} range rather than in the \unit[]{MHz} one are required. The latter is the reason why the devices were not studied in the vortex state, for typically the dynamics lie in the sub-\unit[]{GHz} range \cite{dussaux2010large,locatelli2015efficient,hamadeh2024core,koujok2023resonant}. In the framework of this study, we are interested in rf output of the STNO regardless of its origin, as the aim of this study does not evolve around the characterization of STNO devices.

FIG. \ref{fig2} (b) represents  the STNO output spectrum at fixed $B_{\mathrm{STNO}}$ and $I_{\mathrm{STNO}}$. These values are thereafter held constant at \unit[21]{mT} and \unit[3]{mA}, respectively throughout the study. Only frequencies between \unit[3]{GHz} and \unit[5]{GHz} are shown to match the frequency range of the magnonic circuit, as the latter is limited by the electro-magnet.

\begin{figure}[!ht]
\centering
\includegraphics[width=\columnwidth]{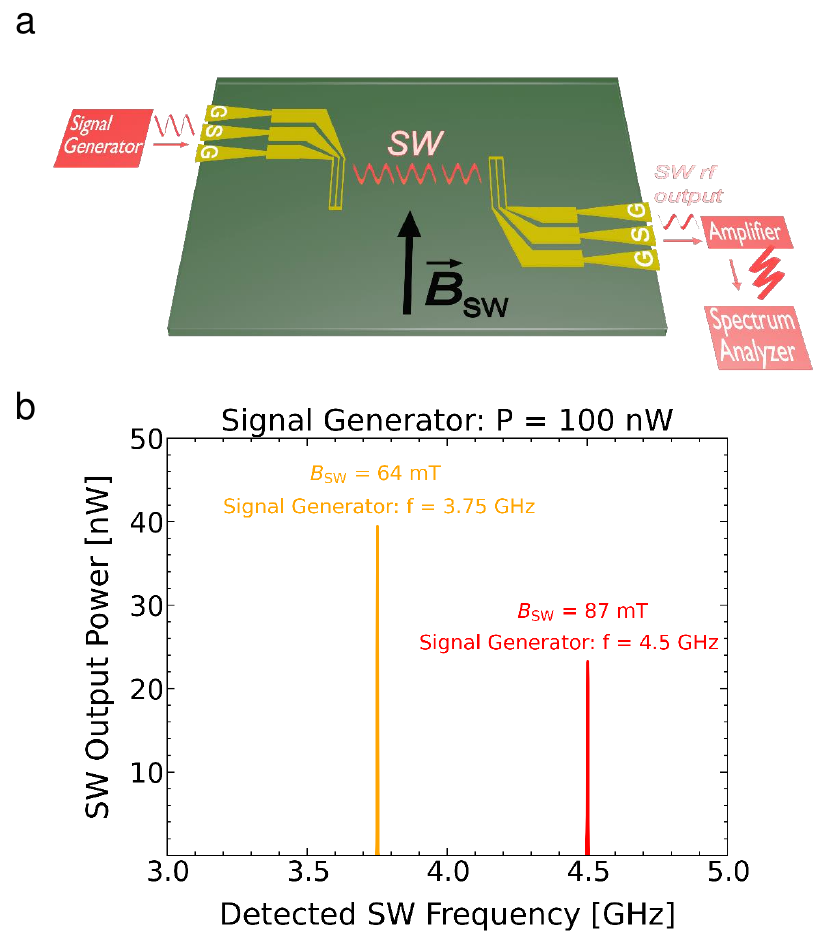}
\caption{(a) Schematic of the magnonic circuit where Coplanar waveguide antennas are used to excite and detect spin waves. The magnetic field $B_{\mathrm{SW}}$ is applied perpendicular to the SW propagation direction. (b) The detected SW signal, as excited by the microwave signal generator, at the resonant Fields $B_{\mathrm{SW}}$.}\label{fig3}
\end{figure}

\begin{figure}[!ht]
\centering
\includegraphics[width=\columnwidth]{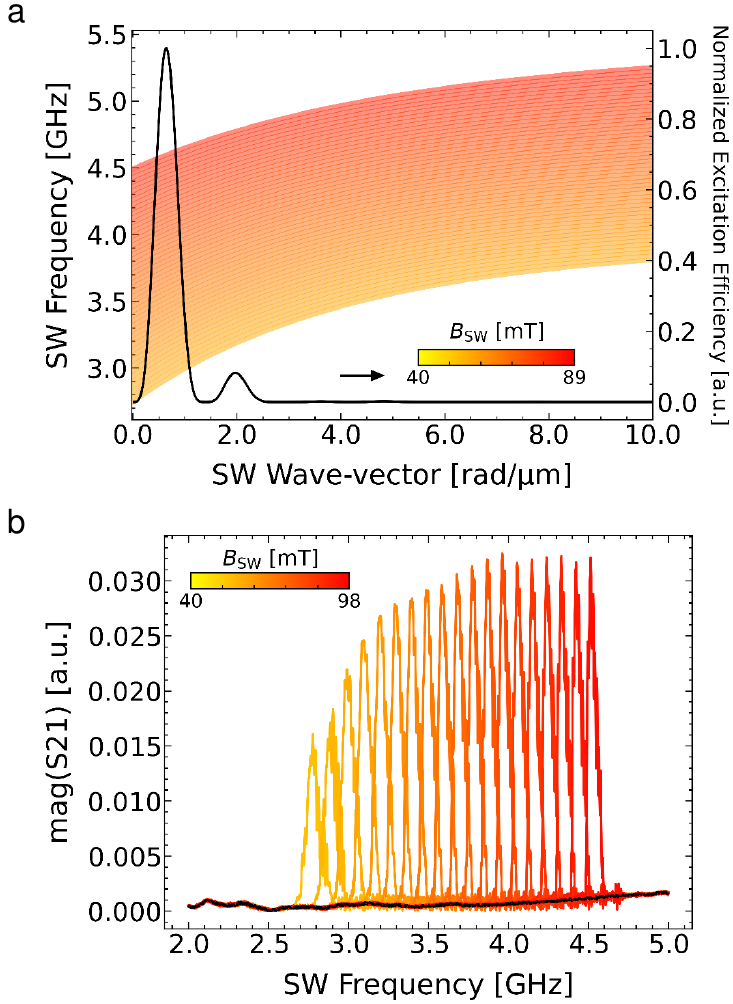}
\caption{(a) Calculated spin wave dispersion curves for the utilized Yttrium-Iron-Garnet chip for fields ranging from \unit[40]{mT} up to \unit[89]{mT}. The antenna's excitation efficiency is shown on the right-hand side. Maximum efficiency lies around wave-vectors close to \unit[0.7]{rad/$\mu$m} (b) Spectra of the forward transmission (component S$_{21}$ of the scattering matrix) as obtained from a vector network analyzer for fields ranging from \unit[40]{mT} up to \unit[98]{mT}. The black line traced at the base of the transmission spectra is the electromagnetic cross-talk between the two CPWs at zero field.}\label{fig4}
\end{figure}

In FIG. \ref{fig3} (a), we present the circuit constituting the SW system. To excite and detect SWs, Coplanar Waveguide (CPW) antennas are fabricated on a \unit[100]{nm} thick YIG chip. The widths of the current carrying lines are $\unit[2]{\mu m}$ for the middle line and $\unit[1]{\mu m}$ the outer lines with a gap of $\unit[2.5]{\mu m}$ in between them. The magnonic circuit is placed between the poles of a second electro-magnet with the generated magnetic field $B_{\mathrm{SW}}$ being perpendicular to the direction of SW propagation. The CPW antennas on both ends, separated by $\unit[30]{\mu m}$, are used to respectively excite and detect SWs. The basic functioning of such a spin wave transducer has already been described in numerous other publications \cite{Bailleul2003PSWS,Vlaminck2010transduction,Kohl2025Loss}. As depicted in FIG. \ref{fig3} (a), for a first characterization, a signal generator connected to one of the CPWs is used to excite SWs at a frequency matching the excitation spectrum of the CPW at the given magnetic field. This excited spin-wave signal can be then detected at the second CPW, from which it is directed to an amplifier, then to a spectrum analyzer. FIG. \ref{fig3} (b) presents the spectrum of the detected SW signal as excited by the microwave signal generator. The signal generator's power is set to the minimum value (P = \unit[100]{nW}). To demonstrate the filtering capability of the magnonic circuit, the magnetic field is swept from \unit[0]{mT} to \unit[88]{mT} after applying a microwave signal to the input CPW at a fixed frequency. To excite/detect SWs at \unit[3.75]{GHz}, a magnetic field $B_{\mathrm{SW}}$ = \unit[64]{mT} is required (see yellow peak in FIG.\ref{fig3}). SW transmission at higher frequencies requires  higher magnetic fields, e.g. a frequency of \unit[4.5]{GHz} can be transmitted using a magnetic field of $B_{\mathrm{SW}}$ = \unit[87]{mT}. The excitation efficiency of the CPW is wavevector specific, which makes it necessary that the applied magnetic field $B_{\mathrm{SW}}$ is adjusted such that the wavevector specificity matches the chosen microwave frequency via the SW dispersion of the material. For this reason, the SW dispersion for this film is calculated whilst highlighting the CPW's excitation spectrum (see FIG. \ref{fig4} (a)). Essentially, the shape of the transmission spectrum depends on the dispersion of the propagating spin waves and the excitation efficiency of the antennas used. The CPW's excitation efficiency spectrum, described by the wave vector spectrum of their magnetic field distribution, is shown in FIG. \ref{fig4} (a). Due to the periodicity of the current distribution, the efficiency shows a maximum at a finite wave vector of \unit[0.7]{rad/$\mu$m}, followed by a second small maximum at higher frequencies. Here, the excitation efficiency describes how efficiently a single antenna excites and detects spin waves of fixed $k$. Using the dispersion relation, the wave vector-dependent efficiency is translated into frequency domain. Further, the group velocity determines the influence of the attenuation of the spin waves during propagation. In FIG. \ref{fig4} (a) the calculated SW dispersion for the \unit[100]{nm} thick YIG chip at different magnetic fields for the Damon-Eshbach geometry (magnetic field applied perpendicular to the direction of propagation of spin waves) is shown. For all magnetic fields shown, the dispersions flatten out towards higher wave vectors. This, in combination with the efficiency, leads to a peak of maximum signal followed by a very weak second peak towards higher frequencies. An exemplary measurement highlighting this behavior for different magnetic fields is shown in FIG. \ref{fig4} (b). For this, a vector network analyzer is used to measure the S$_{21}$ entry of the scattering matrix. This measurement is carried-out for fields ranging from \unit[40]{mT} to \unit[98]{mT}. If the external field is increased, the frequency of the spin waves increases for a fixed k-vector. As a result, increasing the field in FIG. \ref{fig4} (b), the position of the maximum shifts towards higher frequencies. One key point to address here is the reduced transmitted signal amplitude at lower frequencies. These increased insertion losses are related to the increase in impedance mismatch at lower frequencies, where the impedance of the magnonic device is significantly lower than the 50 $\Omega$ impendances of the connected devices. For this reason, we apply a sufficiently large in-plane field to operate the STNO in the uniform mode regime at frequencies that match those of the magnonic system in the range with small insertion losses (\unit[3]{GHz} and above). In the in-plane oscillating regime, the lowest frequency mode is on the order of \unit[]{GHz}, and hence can yield a better SW frequency matching for the studied magnonic system. For detailed information and quantitative measurements on the characterization of the SW excitation efficiencies for this film but also other thicknesses, CPW antenna dimensions and impedance, in addition to determining the loss channels and the energy dissipated through those channels, refer to \cite{kohl2025identification}.

\begin{figure*}[!ht]
\centering
\includegraphics[width=0.8\textwidth]{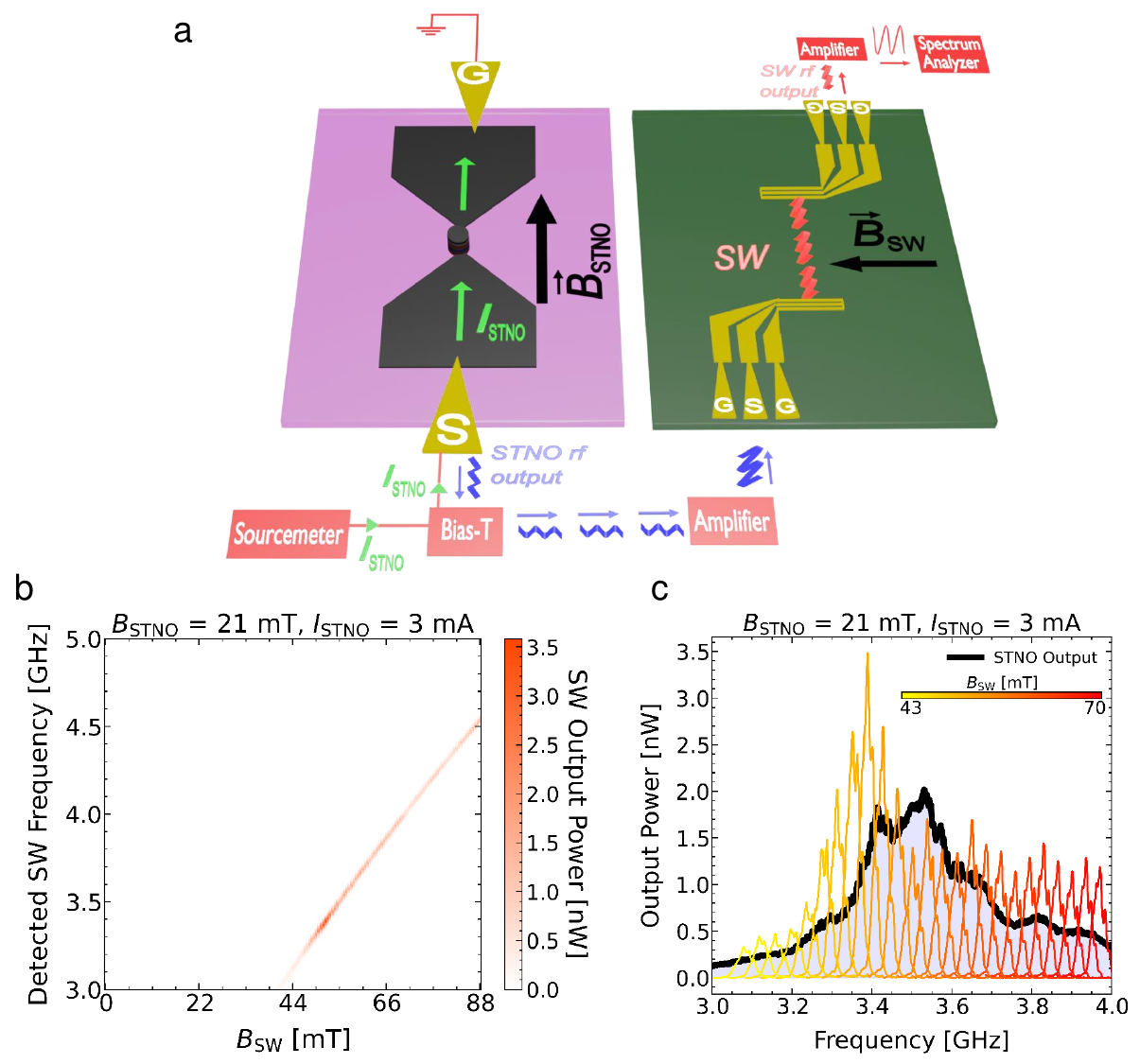}
\caption{(a) The proposed concept of the magnonic filter consisting of the source (STNO circuit) and the filtering device (SW circuit). (b) 2D color-plot of the frequency dependence of the SW's output signal on the applied magnetic field $B_{\mathrm{SW}}$ at fixed STNO conditions: $B_{\mathrm{SW}}$ = \unit[21]{mT} and $I_{\mathrm{STNO}}$ = \unit[3]{mA}. (c) Signals filtered by the SW layer for different $B_{\mathrm{SW}}$ values ranging from \unit[43]{mT} to \unit[70]{mT} in reference to the broad input signal from STNO (black curve).}\label{fig5}
\end{figure*}

FIG. \ref{fig5} (a) is a schematic of the merged system combining the STNO and the SW circuits. The same aforementioned approach regarding RF signal generation is followed for the STNO, however its output is now plugged-in as an input into the magnonic system. Simply put, the STNO hereby and forth replaces the signal generator used to excite SWs (see FIG. \ref{fig3}). As for the remainder of the SW circuit, it is preserved as the SWs now excited by the STNO's input are detected by means of the second CPW antenna. The RF signal is then amplified and passed onto the spectrum analyzer. Thus, the entire detected signal has been transmitted through the SW domain. First, we fix the conditions at the STNO with $B_{\mathrm{STNO}}$ = \unit[21]{mT} and $I_{\mathrm{STNO}}$ = \unit[3]{mA} (the exact same modes as in FIG. \ref{fig2} (b) are excited). Then, $B_{\mathrm{SW}}$ is swept and the frequency variation versus $B_{\mathrm{SW}}$ is presented in FIG. \ref{fig5} (b). A similar behavior as previously exhibited in FIG. \ref{fig3} (b) is observed even though the signal used to excite the SWs is much broader with a bandwidth of approximately \unit[1]{GHz}. This means that the SW layer can output narrowband signals regardless of the specificity and the quality of the source used to generate these signals. In other words, it is possible to selectively filter a (deteriorated) input signal just by tuning $B_{\mathrm{SW}}$. In FIG. \ref{fig5} (c), the filtering of the broad input signal (in black) is clearly presented as a function of $B_{\mathrm{SW}}$. Filtering is indeed possible for frequencies outside the presented range as well, however those happen to either lie outside the amplification range and/or the accessibility of the electro-magnet at the SW layer. Within the studied range of $B_{\mathrm{SW}}$ in FIG. \ref{fig5} (c), only a few filtered signals are shown for presentation purposes. Setting the appropriate field ($B_{\mathrm{SW}}$), the desired pass band can be chosen. The selectivity of the filtering (i.e. the bandwidth of the filtered signal) depends on the dispersion relation in the YIG and the wavevector range excited by the CPW, which is about \unit[100]{MHz} in this case. Additionally, in \textbf{section \RomanNumeralCaps{1}} of \textbf{Supplementary Material}, we evidence the independence of this SW filtering on the conditions at the STNO. We show that changing the field solely on the magnonic part (($B_{\mathrm{SW}}$)) is what defines the frequency at which the broad input is filtered. As can be noticed from the filtered spectra in FIG. \ref{fig5} (c), the envelope formed by the maxima of the SW signal does not coincide with those of the STNO input. The reason behind this is impedance mismatch, as impedance on the SW system is frequency dependent, and can thus be further/closer to that of the STNO at different frequencies. This results in a frequency dependent efficiency of power transfer from the source (STNO) into the SW system. At frequencies where the impedance mismatch is larger, a more significant portion of the STNO's input into the SW system is reflected back, hence leading to weaker transmitted SW signal. The impedance mismatch hence causes a change in the ratio of the transmitted power to the output power (by the STNO), which results in a different trend if one is to consider the S21 parameter in this case (compared to FIG. \ref{fig4} (b) where a VNA was used).

Thus, we are able to demonstrate the detection/emission of SWs excited by a STNO, and find means of use for broad output STNOs. In addition, we suggested a direct application to employ the proposed device-concept as an RF source with an integrated magnonic filter that offers both frequency tunability and selectivity. This introduces an approach to RF signal processing where a nanoscaled spintronic oscillator can directly drive a magnonic circuit, opening possibilities for compact, tunable microwave filters in magnon-based computing or sensing applications. Our work constitutes a promising first step in that direction with room for further development. Optimization techniques should address better impedance matching between the STNO and the magnonic system, increasing the output power of the STNOs to avoid the need of amplification. A way to realize this would be by reconfiguring or re-engineering the material stack accordingly, in addition to employing a larger number of coupled oscillators in the form of an array to generate larger output powers, and exhibit a higher resiliency to external conditions \cite{qu2015phase,lebrun2017mutual}. On the other hand, it is also possible to optimize the CPW designs to get a more selective excitation with a narrower linewidth in wave-vector space. Ultimately, the long-term goal is to fabricate the STNOs directly on the SW chip. This eliminates the need for electrical connection between the circuits as SWs are then excited by the dynamic magnetic stray fields generated by the STNOs. The system will use an on-chip STNO as the source for an on-chip magnonic filter. This integration promises benefits for RF systems, such as eliminating the need for a high quality LC filter or external synthesizer in the accessible frequency range, for example). Such a system has been previously investigated via micromagnetic simulations in \cite{hamadeh2022hybrid}. Additionally, the STNO system can be designed to operate at zero bias field \cite{hamadeh2024core}. For the magnonic system, integrated on-chip permanent magnets can be used to manipulate and tune SW dispersion \cite{cocconcelli2024tuning,cocconcelli2025standalone,kohl2025identification}. In \textbf{section \RomanNumeralCaps{2}} of \textbf{Supplementary Material}, we present a qualitative energy-efficient approach to manipulating the filtered signal's frequency, whereby we shift this frequency by means of a small permanent block magnet suspended above the YIG delay line. The altitude of the latter above the magnetic material influences the frequency at which SWs are emitted/detected via a tunable additional field component.


In conclusion, we demonstrate a hybrid system where a spintronic nano-oscillator (with a broad \unit[]{GHz} output) directly excites a magnonic circuit. By tuning the magnetic field on the magnonic side, we achieve selective filtering of the broad input—delivering a narrow-band SW output at a chosen frequency, irrespective of the input’s linewidth. This tunable magnonic filter concept could allow detection and cleanup of deteriorated RF signals in a field-programmable manner, reducing reliance on charge-based components.This work provides the  experimental proof that a STNO can drive a magnonic filter, paving the way for integrated magnonic-spintronic circuits for microwave signal processing. 

\section*{Supplementary Material}
For an additional measurement demonstrating filtering achieved by sweeping $I_{\mathrm{STNO}}$ instead of $B_{\mathrm{SW}}$, and evidencing the importance of $B_{\mathrm{SW}}$ in tuning the filtered frequencies, please refer to the \textbf{Supplementary Experimental Measurement \RomanNumeralCaps{1}}.\\
For an additional measurement demonstrating the tunability of the filtered signal by means of a permanent block magnet, please refer to the \textbf{Supplementary Experimental Measurement \RomanNumeralCaps{2}}

\section*{Acknowledgements}
This work has been  supported  by the European Research Council within the Starting Grant No. 101042439 "CoSpiN", the Deutsche Forschungsgemeinschaft (DFG, German Research Foundation) - TRR 173 - 268565370" (project B01) and  the EU Horizon Europe research and innovation program within the project “MandMEMS” (Grant No. 101070536).

\section*{References}

\bibliography{references}

\end{document}


\title[Article Title]{Hybrid magnonic–spintronic system for tunable broadband signal filtering and microwave generation}

\keywords{}

\maketitle

\section*{Supplementary Experimental Measurement \RomanNumeralCaps{1}}

\begin{figure}[!ht]
\centering
\includegraphics[width=0.8\textwidth]{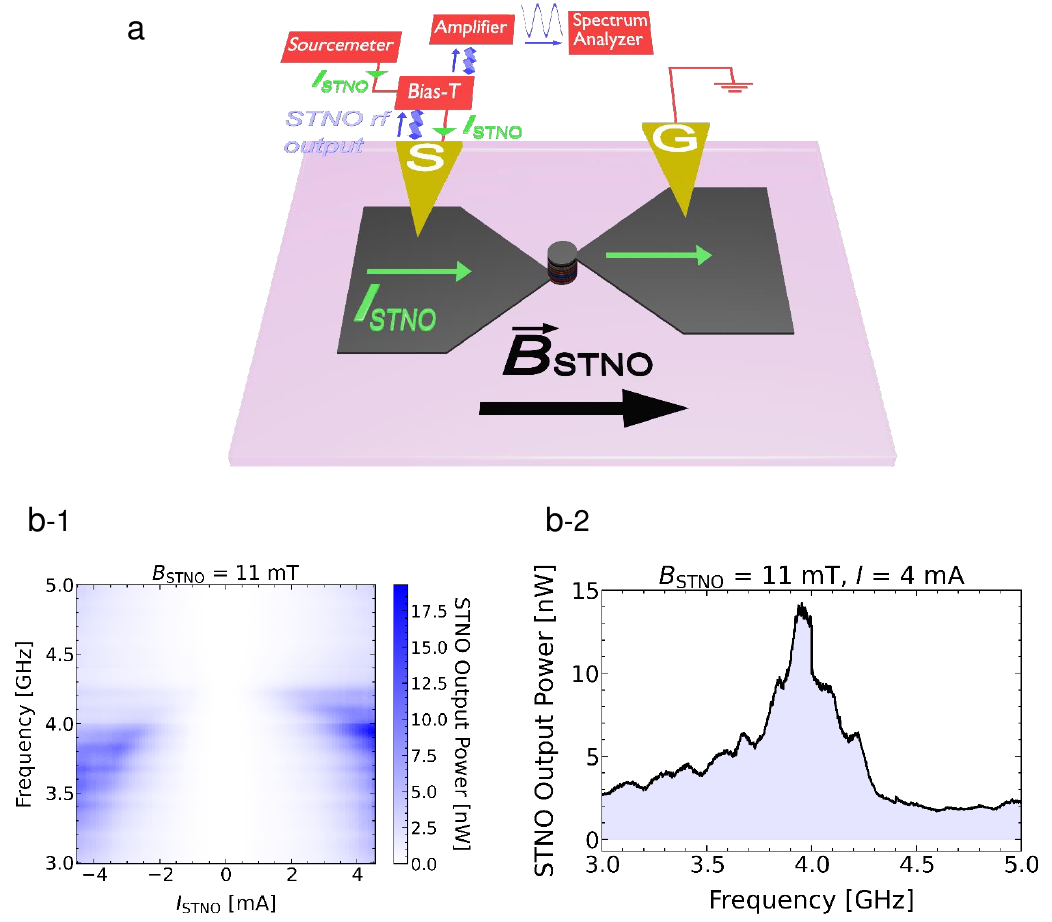}
\caption{(a) The schematic of the STNO layer as presented earlier in the main text. (b-1) 2D color-plot showing the STNO's frequency change as a function of the swept DC current ($I_{\mathrm{STNO}}$) at $B_{\mathrm{STNO}}$ = \unit[11]{mT}. (b-2) Power spectrum of displaying the STNO's auto-oscillating mode(s) at $B_{\mathrm{STNO}}$ = \unit[11]{mT} and $I_{\mathrm{STNO}}$ = \unit[4]{mA}.}\label{fig1}
\end{figure}

In this section, supplementary measurements are carried-out on a different STNO of the same dimensions as the one used in the main text. As such, the same circuit previously used to measure the RF output of the STNO is deployed (see FIG. \ref{fig1} (a)). 

Alternatively, we hereby sweep $I_{\mathrm{STNO}}$ instead of $B_{\mathrm{STNO}}$. FIG. \ref{fig1} (b-1) is a 2D color-plot of the frequency dependence of the STNO on DC current $I_{\mathrm{STNO}}$. After saturating the magnetic layers at high fields above \unit[500]{mT}, $I_{\mathrm{STNO}}$ is swept between \unit[-4.5]{mA} and \unit[4.5]{mA} at constant field $B_{\mathrm{STNO}}$ = \unit[11]{mT}. Two distinct broad excitations are observed at the two opposite ends of the color-plot, with maximum output power achieved for $I_{\mathrm{STNO}}$ ranging between \unit[4]{mA} and \unit[4.5]{mA}. This clearly shows that the STNO is not in the normal auto-oscillatory regime where damping is compensated for one current direction and enhanced for the opposite one. As such, the detected signal is, as mentioned in the main text, due to noise amplification.
Shown in FIG. \ref{fig1} (b-2) is a spectrum where an excitation with high power is adopted with $I_{\mathrm{STNO}}$ fixed at \unit[4]{mA}. FIG. \ref{fig4} (b-2) represents the output signal as generated by the STNO, which in turn will be input into the SW circuit.

\begin{figure}[!ht]
\centering
\includegraphics[width=0.8\textwidth]{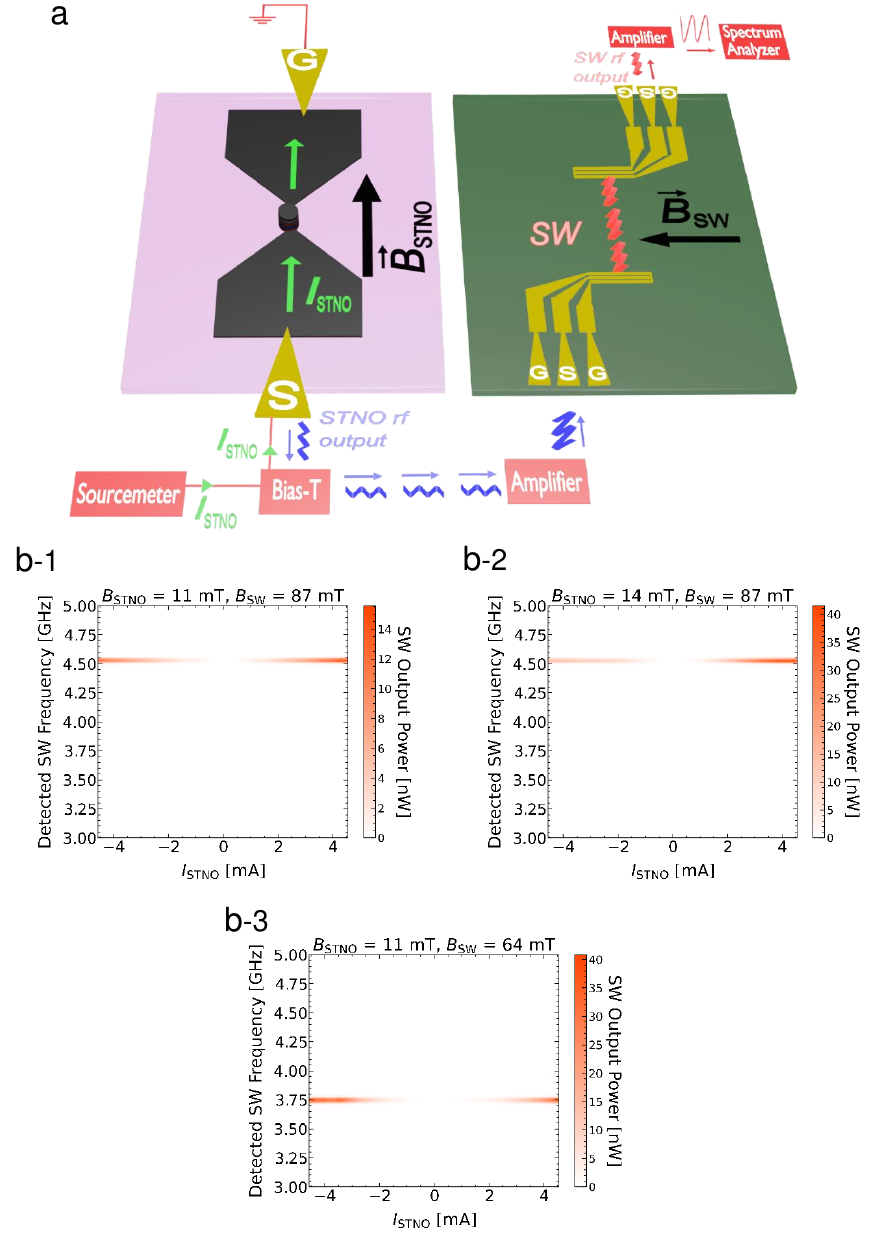}
\caption{(a) The schematic of the combined layers as presented earlier in the main text. (b) Signal filtering as a function of the swept DC current $I_{\mathrm{STNO}}$ for different field conditions (1) $B_{\mathrm{STNO}}$ = \unit[11]{mT}, $B_{\mathrm{SW}}$ = \unit[87]{mT} (2) $B_{\mathrm{STNO}}$ = \unit[14]{mT}, $B_{\mathrm{SW}}$ = \unit[87]{mT} (3) $B_{\mathrm{STNO}}$ = \unit[11]{mT}, $B_{\mathrm{SW}}$ = \unit[64]{mT}.}\label{fig2}
\end{figure}

In FIG. \ref{fig2} (a), the same combined setup scoping the spintronics circuit and the magnonic one is presented for clarification. The STNO's RF output is now used to generate SWs at the magnonic circuit. FIG. \ref{fig2} (b) shows the achieved filtering of STNO's input as a function of the same current sweep presented in FIG. \ref{fig1} (b-1). First, $I_{\mathrm{STNO}}$ is swept between \unit[-4.5]{mA} and \unit[4.5]{mA}, and the magnetic fields at both setups are fixed with $B_{\mathrm{STNO}}$ = \unit[11]{mT} and $B_{\mathrm{SW}}$ = \unit[87]{mT} (FIG. \ref{fig2} (b-1)). At the spectrum analyzer, the detected signal has a frequency of nearly \unit[4.55]{GHz} for all values of $I_{\mathrm{STNO}}$. Next, we increase $B_{\mathrm{STNO}}$ to \unit[14]{mT} whilst keeping $B_{\mathrm{SW}}$ unchanged (FIG. \ref{fig2} (b-2)). The same detected frequency persists. The lower power of the detected signal is a result of the change in the STNO's output power for a different magnetic field. Nevertheless, the persistence of the same detected frequency for a different field value at the spintronic setup agrees with the results we presented in the main text. In other words, the frequency of the filtered signal is solely determined by $B_{\mathrm{SW}}$. To evidence this for the dc current sweep, we now present the case where we keep $B_{\mathrm{STNO}}$ at \unit[11]{mT}, as it was previously in FIG. \ref{fig2} (b-1), and only change $B_{\mathrm{SW}}$ from \unit[87]{mT} to \unit[64]{mT} (see FIG. \ref{fig2} (b-3)). Indeed, the filtered signal's frequency now drops to \unit[3.75]{GHz} which again proves that selective signal filtering can be achieved by tuning $B_{\mathrm{SW}}$.

\newpage

\section*{Supplementary Experimental Measurement \RomanNumeralCaps{2}}

\begin{figure}[!ht]
\centering
\includegraphics[width=0.8\textwidth]{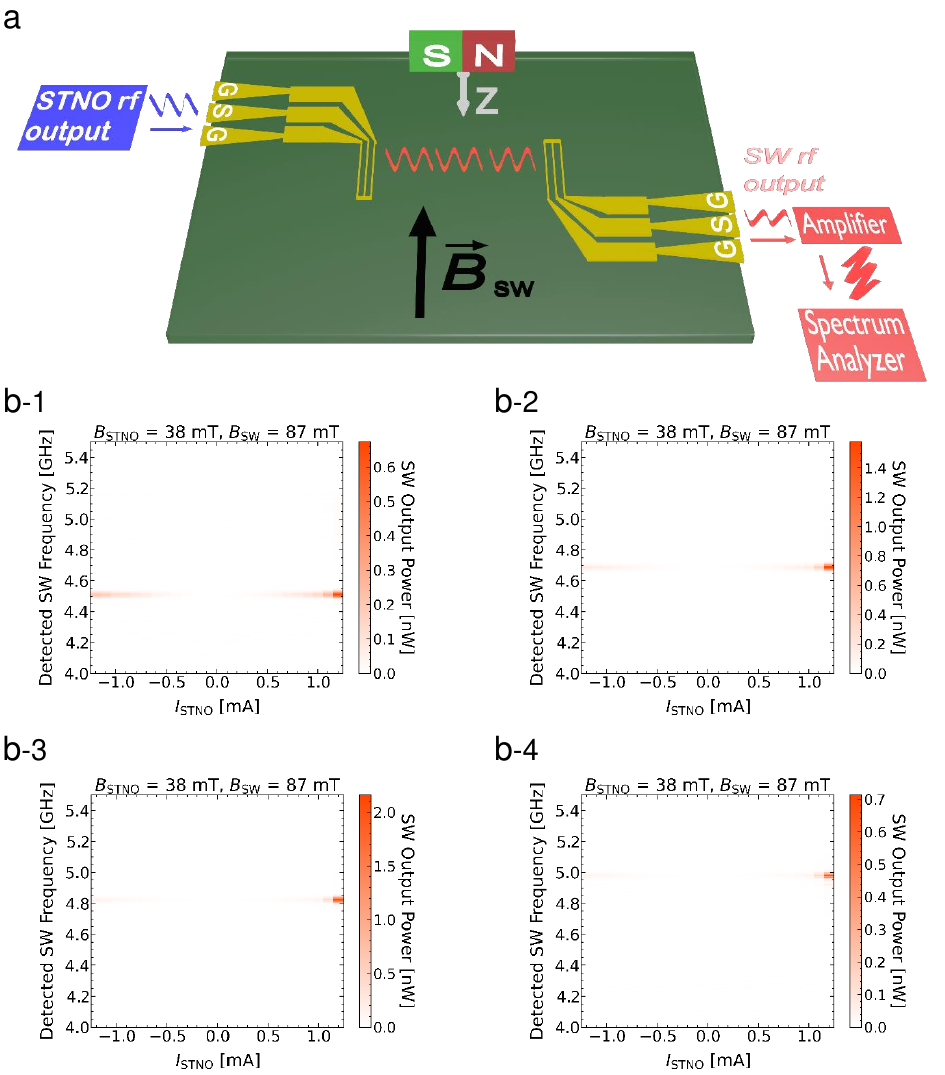}
\caption{(a) Simplified schematic of the setup used to manipulate SW frequencies. The height of a Neodymium block magnet is varied to shift the SW dispersion. (b-1,2,3,4) color-plots of the magnet-tuned signal filtering for same field conditions and current sweeps.}\label{fig3}
\end{figure}

Here, we present additional supplementary measurements on a STNO that has a diameter of \unit[250]{nm} (also works at the diameters used earlier, but those MTJs degraded because of the measurements), whereby we introduce a tunable energy-efficient means to manipulate SW frequencies by utilizing a translatable small permanent block magnet. For this, the same exact procedure for microwave generation presented in the main text of the manuscript is followed.

FIG. \ref{fig3} (a) is a simplified schematic focusing on the method used to achieve the aforementioned energy-efficient tunability. As the spintronic has not been changed, we omit the circuit for clarity purposes, and only display it as a microwave source (blue box on the left). As clarified before, the STNO's input into the SW layer excites SWs at a frequency set by $B_{\mathrm{SW}}$. At a proximity from the CPW antennas, we introduce a permanent block magnet that can generate a magnetic field of nearly \unit[90]{mT}. The magnet has the following dimensions: 5$\times$1.5$\times$1 mm$^3$, and is magnetized height-wise. The magnet is however rotated by 90 degrees such that its short face opposes the structure, and hence introduces an in-plane component instead of an out-of-plane one. At the microscopic picture, this magnet is still considered huge, however the idea here is to provide a proof-of-concept for an additional versatile frequency tunability via mechanical means in such systems. The use of smaller magnets is indeed needed for practical applications, however, as long as the stray field generated by the magnet is in good proximity with the magnetic structure, changes in the SW frequency can be achieved.

\begin{figure}[!ht]
\centering
\includegraphics[width=0.8\textwidth]{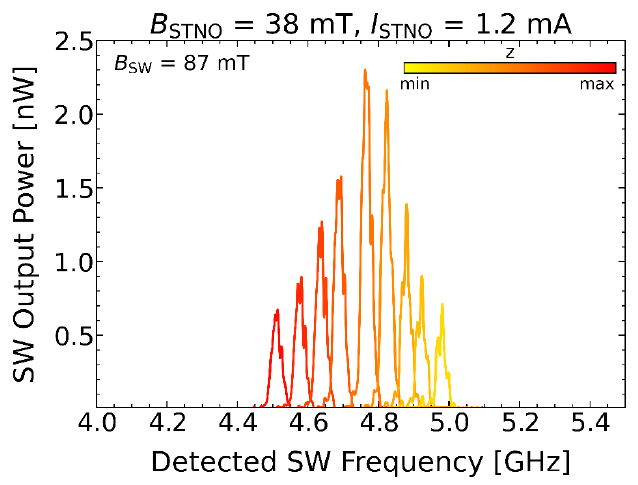}
\caption{The varying frequency and power of the SW output signal as a function of the magnet's altitude z.}\label{fig4}
\end{figure}

In FIG. \ref{fig3} (b), we present four different filtered spectra for same $B_{\mathrm{STNO}}$ and $B_{\mathrm{SW}}$ conditions whilst sweeping $I_{\mathrm{STNO}}$ between \unit[-1.2]{mA} and \unit[1.2]{mA}. Starting by FIG. \ref{fig3} (b-1), the block magnet is held at a relatively large distance (z) above the sample. This can be intuitively understood simply by comparing FIG. \ref{fig3} (b-1) with FIG. \ref{fig2} (b-1). As can be seen, the SW frequency is practically the same which means that the effects of the block magnet are null in this case. As z is decreased and the magnet gets closer to the sample, a positive shift in the SW's output frequency is recorded (FIG. \ref{fig3} (b-2)). Decreasing z even further, the frequency continues the increasing trend (FIG. \ref{fig3} (b-3,4)). The frequency difference between the magnet being at the largest value of z and at the smallest one is nearly \unit[500]{MHz}. As a rough estimation, the magnetic field that needs to be added to the system as to induce this change in frequency is around \unit[17]{mT} (this estimation was obtained from a separate measurement). In FIG. \ref{fig4}, we present the power spectra of the filtered signal for fixed $B_{\mathrm{STNO}}$, $I_{\mathrm{STNO}}$ and $B_{\mathrm{SW}}$ conditions. By referring to the DC current sweep from FIG. \ref{fig3} (b), $I_{\mathrm{STNO}}$ is fixed at \unit[1.2]{mA} since the signal power is maximal for each case. Here, more steps in between the interval of the \unit[500]{MHz} shift are presented. Indeed, decreasing z gradually shifts the SW frequency towards higher values. The observed change in the power of the transmitted signal (as can be inferred from FIG. \ref{fig3} (b) as well) can be traced back to the block-magnet induced inhomogeneity of the magnetic field in the vicinity of the generated SW, in addition to impedance mismatches between both circuits. This however can be optimized further through fabrication protocols aiming at engineering better impedance matching, and the possibility of utilizing a smaller permanent magnet which can be integrated on-chip for instance.